\documentclass[12pt]{article}
\usepackage[usenames,dvipsnames,svgnames,table]{xcolor}
\usepackage{amsmath,lscape,amsfonts,amssymb,amsthm,verbatim,lscape}
\usepackage[figuresright]{rotating}
\usepackage{natbib}
\usepackage{url}
\usepackage{hyperref}
\hypersetup{colorlinks,
citecolor=black,
filecolor=black,
linkcolor=black,
urlcolor=black,
pdftex}

\usepackage{booktabs}

\usepackage{colortbl}
\definecolor{gray}{rgb}{0.8,0.8,0.8}

\usepackage{times}
\usepackage{blindtext}
\usepackage[font={footnotesize,it}]{caption}

\usepackage{sectsty}  
\subsectionfont{\normalsize\bf}
\sectionfont{\large\bf}

\def\atil{{\widetilde\a}}

\newcommand{\bea}{\begin{eqnarray}}
\newcommand{\eea}{\end{eqnarray}}
\def\bbf{\boldsymbol{\beta}}

\def\gbf{\boldsymbol{\gamma}}

\def\thbf{\boldsymbol{\theta}}

\def\Debf{\boldsymbol{\Delta}}

\def\a{\mathbf{a}}
\def\g{\mathbf{g}}

\def\s{\mathbf{s}}

\def\V{\mathbf{V}}
\def\bfD{\mathbf{H}}

\def\R{\mathbf{R}}

\def\M{\mathbf{J}}

\def\W{\mathbf{W}}
\def\x{\mathbf{x}}
\def\X{\mathbf{X}}

\def\y{\mathbf{y}}

\def\b{\beta}

\def\O{{\boldsymbol\Omega}}
\def\bfPsi{{\mathbf\Psi}}

\def\p{\partial}

\begin{document}

\def\spacingset#1{\renewcommand{\baselinestretch}%
{#1}\small\normalsize} \spacingset{1}

\title{Efficient and feasible inference for high-dimensional\\ normal copula regression models}

\author{Aristidis~K.~Nikoloulopoulos \footnote{\href{mailto:a.nikoloulopoulos@uea.ac.uk}{a.nikoloulopoulos@uea.ac.uk},  School of Computing Sciences, University of East Anglia, Norwich NR47TJ, U.K.} }

\date{}

\maketitle

\begin{abstract}
\spacingset{1.9}

 \noindent The composite  likelihood (CL) is amongst the computational methods used for the estimation of high-dimensional multivariate normal (MVN) 
copula models with discrete responses. Its computational advantage,  as a surrogate likelihood method, is that is based on the independence likelihood for the univariate regression and non-regression parameters and pairwise likelihood for the correlation parameters, but the efficiency of estimating the univariate regression and non-regression parameters can be  low. 
For a high-dimensional discrete response, we propose weighted versions of the composite likelihood estimating equations and an iterative approach to determine good weight matrices.   
The general methodology is applied to the MVN copula with univariate ordinal regressions as the marginals.  Efficiency calculations show that our method is nearly as efficient as the maximum likelihood for fully specified  MVN
copula models. 
Illustrations include simulations and real data applications regarding longitudinal (low-dimensional)  and time (high-dimensional) series ordinal response data with covariates and it is shown that there is a substantial gain in efficiency via the weighted CL method.
 
\end{abstract}

\spacingset{1.9} 

\section{Introduction}
  The multivariate normal (MVN) copula with discrete margins,
has been in use for a considerable length of time, e.g.,  \cite{joe97}, and much earlier in the biostatistics \citep{Ashford&Sowden1970}, psychometrics \citep{Muthen1978}, and econometrics \citep{Hausman&Wise1978} literature.  It is usually known as a multivariate or multinomial probit model.
The multivariate probit model is a simple example of the MVN copula with univariate probit regressions as the marginals.
The use of the MVN copula with logistic regression (or Poisson or negative binomial or ordinal regression) is just a special case of the general theory of dependence modelling with copulas (e.g., \citealt{joe2014}).

The MVN copula is generated by the MVN distribution and thus inherits the useful properties of the latter. Therefore, the MVN copula
allows a wide range of dependence and overcomes the drawback of limited dependence inherent in simple parametric
families of copulas \citep{Nikoloulopoulos2013a,nikoloulopoulos13b}. When the univariate margins are regression models for discrete response, then copula models can be more difficult to discriminate \citep[page 242]{joe2014}. Hence, the MVN  copula with discrete margins or discretized MVN model 
does not only provide a wide range of flexible dependence but also approximates other copula models.

Nevertheless,  implementation of the MVN copula for discrete data  is not easy, because the MVN distribution as a latent model for discrete response requires  rectangle probabilities based on high-dimensional integrations or their approximations  \citep{Nikoloulopoulos&karlis07FNM,
panagiotelis&czado&joe12}
unless the correlation matrix is positive exchangeable  or has an 1-factor structure \citep{Johnson&Kotz72}.
Hence, \cite{joe97} and  \cite{song07} 
restricted on low dimensional regression modelling of  dependent discrete  data using the MVN copula.

\cite{nikoloulopoulos13b,nikoloulopoulos2015a}
and \cite{Masarotto&Varin12} proposed efficient simulated likelihood methods that can be used for estimation of MVN copula discrete regression models in higher dimensions, occurring with time series, spatial data, longer longitudinal
studies, but there is an issue of computational burden as the dimension $d$ and the sample size $n$ increase. 
This is also the case for the efficient Bayesian data augmentation method of \cite{pitetal2006}  or the data augmentation together with a parameter expansion approach of \cite{Murray-etal-2013} as  the number of latent variables is of the same size as the data, i.e., a matrix of size $n \times d$ (e.g., \citealt{panagiotelis&czado&joe12}).

 \cite{zhao&joe05} proposed composite likelihood
estimating equations
to overcome the computational issues at the maximization routines for the MVN copula in a high-dimensional context by using the independence likelihood for the marginal parameters and pairwise likelihood for the correlation parameters. Composite likelihood is  a surrogate likelihood which leads  to unbiased estimating equations \citep{varin08,Varin-etal2011} obtained by the derivatives of the composite univariate and bivariate log-likelihoods.  As the estimation of univariate parameters  ignores the dependence,  the efficiency of  estimating  the univariate regression and non-regression parameters is low.

To improve the efficiency of the composite likelihood method on estimating the MVN copula with discrete margins, we propose weighted versions of the composite likelihood estimating equations and an  iterative approach to determine good weight matrices.  Based on the matrix version of the Cauchy-Schwarz inequality \citep{chaganty97b}, we  determine the optimal weights for which the asymptotic efficiency of the proposed  estimates  with these weights is close to the asymptotic efficiency of the maximum likelihood (ML) estimates. Our intent is to develop  an efficient estimation method for MVN copula regression models for discrete responses that 
can be used for the regression analysis of high-dimensional discrete response data with covariates observed in
time series or spatial statistics.

The remainder of the paper proceeds as follows. Section \ref{theory-sec} introduces the general theory of weighted versions of the composite likelihood estimating equations and gives the details for the MVN copula with ordinal regressions as the marginals. Section \ref{asymptotics-sec} studies asymptotic efficiency of our method as compared to the `gold standard' ML method. Section \ref{sim-sec} studies the small-sample efficiency of the weighted composite score functions in both  low- and high-dimension. Section \ref{sec-app} presents  two applications of our methodology to analyze longitudinal (low-dimensional) and time (high-dimensional) series ordinal response data. We conclude this article with some
discussion.

\section{\label{theory-sec} Weighted versions of the composite likelihood estimating equations}

To illustrate the method of the weighted  versions of the composite likelihood estimating equations  to estimate the MVN copula parameters concretely, we use  univariate ordinal probit/logit regressions as the marginals. The resulting multivariate discrete distribution is the  multivariate ordinal probit/logit model.

Suppose that the data are $(y_{ij},\x_{ij})$, $j=1,\ldots,d$, $i=1,\ldots,n$,
where $i$ is an index for individuals or clusters and $j$ is an index for
the repeated measurements or within cluster measurements.
The  MVN copula model  with ordinal probit regressions as the marginals has the following cumulative distribution function (cdf):
 $$F(y_{1}, \dots, y_{d}; \nu_1,\ldots,\nu_d,\gbf,\R)=\Phi_d\left(\Phi^{-1}[F_1(y_{1};\nu_{1},\gbf)],\ldots,
 \Phi^{-1}[F_1(y_{d};\nu_{d},\gbf)];\R\right),$$
where $\Phi_d$ denotes the standard MVN distribution function with correlation
matrix $\R=(\rho_{jk}: 1\le j<k\le d)$, $\Phi$ is cdf of the univariate standard normal,
and $F_1(y;\nu,\gbf)=\mathcal{F}(\alpha_y+\nu)$ is the univariate
cdf for $Y$, where $\nu=\x^T\bbf$ is a function of $\x$
and the $p$-dimensional regression vector $\bbf$, and $\gbf=(\alpha_1,\ldots,\alpha_{K-1})$ is the $q$-dimensional vector of the univariate cutpoints ($q=K-1$). Note that $\mathcal{F}$ normal leads to the probit model and $\mathcal{F}$ logistic
leads to the cumulative logit model for ordinal response \citep[Section 3.3.2]{agresti2010}.

The MVN copula lacks a closed form cdf; hence implementation of the discretized MVN is feasible, but not easy, because the MVN distribution as a latent model for discrete response requires rectangle probabilities of the form
\begin{equation}\label{MVNrectangle}
f_d(\y_{i})=\int_{\Phi^{-1}[F_{1}(y_{i1}-1;\nu_{i1})]}^{\Phi^{-1}[F_{1}(y_{i1};\nu_{i1})]}
\cdots \int_{\Phi^{-1}[F_{1}(y_{id}-1;\nu_{id})]}^{\Phi^{-1}[F_{1}(y_{id};\nu_{id})]}  \phi_d(z_1,\ldots,z_d;\R) dz_1\cdots dz_d
\end{equation}
where $\phi_d$ denotes the standard $d$-variate normal density  with correlation matrix $\R$. When the joint probability is too difficult to compute, as in the case of the the discretized MVN model, composite likelihood is a good alternative \citep{varin08,Varin-etal2011}.

 \cite{zhao&joe05} proposed the CL
method to overcome the computational issues at the maximization routines for the MVN copula in a high-dimensional context. Estimation of the model parameters
can be approached by solving the estimating equations obtained by the derivatives of the sums of univariate and bivariate log-likelihoods.

The sum of univariate 
log-likelihoods is
\begin{equation}\label{L1}
L_1= \sum_{i=1}^n\sum_{j=1}^d\, \log f_1(y_{ij};\nu_{ij},\gbf)=\sum_{i=1}^n\sum_{j=1}^d\,
 \ell_1(\nu_{ij},\gbf,\, y_{ij}),
\end{equation}
where $f_1(y;\nu,\gbf)=\mathcal{F}(\alpha_{y}+\nu)-\mathcal{F}(\alpha_{y-1}+\nu)$ and  $\ell_1(\cdot)=\log \, f_1(\cdot)$. The score equations for $\bbf$ and $\gbf$ are
\bea\label{gs3}
\begin{pmatrix}\frac{\p L_1}{\p \bbf}\\\frac{\p L_1}{\p \gbf}\end{pmatrix}=
\sum_{i=1}^n\sum_{j=1}^d\, \begin{pmatrix}
\x_{ij} & {\bf 0} \cr
{\bf 0} & {\bf I}_q \end{pmatrix} \begin{pmatrix}
\frac{\p \ell_1(\nu_{ij},\gbf,\, y_{ij})}{\p\nu_{ij}} \cr
\frac{\p\ell_1(\nu_{ij},\gbf,\, y_{ij})}
 {\p\gbf} \end{pmatrix}=
\sum_{i=1}^n\sum_{j=1}^d\,
\begin{pmatrix}
\x_{ij}\mathbf{1}_q \cr
 {\bf I}_q \end{pmatrix}
 \frac{\p \ell_{1ij}(\gbf_{ij}, y_{ij})}{\p\gbf_{ij}}
={\bf 0},
 \eea
where $\gbf_{ij}=(\alpha_1+\nu_{ij},\ldots,\alpha_{K-1}+\nu_{ij})=(\gamma_{ij1},\ldots,\gamma_{ij,K-1})$, $\ell_{1ij}(\cdot)=\log f_{1ij}(\cdot), f_{1ij}(\gbf_{ij}, y)=\mathcal{F}(\gamma_{ijy})-\mathcal{F}(\gamma_{ij,y-1})$,   ${\bf I}_q$ is an identity matrix of
dimension $q$ and $\mathbf{1}_q$ is a unit vector of size $q$.
Let $\X_{ij}^T=\begin{pmatrix}
\x_{ij}\mathbf{1}_q \cr
 {\bf I}_q \end{pmatrix}$
and   $ \s_{ij}^{(1)}(\a)=\frac{\p \ell_{1ij}(\gbf_{ij}, y_{ij})}{\p\gbf_{ij}}$, where $\a^\top=(\bbf^\top,\gbf^\top)$ is the
column vector of all  $r=p+q$ univariate parameters.
The score equations in  (\ref{gs3}) can be written as
 \begin{equation}\label{independent score equations}
\g_1=\g_1(\a)=\frac{\p L_1}{\p \a}=  \sum_{i=1}^n\sum_{j=1}^d\X_{ij}^\top\;\s_{ij}^{(1)}(\a)=
\sum_{i=1}^n\X_i^\top\;\s_i^{(1)}(\a)={\bf 0},
\end{equation}
where
$\X_{i}^\top=(\X_{i1}^\top, \ldots, \X_{id}^\top)$ and $\s_i^{(1)\top}(\a)=
(\s_{i1}^{(1)\top}(\a),\ldots,\s_{id}^{(1)\top}(\a) )$.
The vectors  $\s_{ij}^{(1)}(\a)$ and $\s_i^{(1)}(\a)$ have dimensions $q$ and
$dq$ respectively.
The dimensions of $\X_{ij}$ and $\X_i$ are $q\times r$ and  $dq \times r$
respectively.

The  sum of bivariate log-likelihoods is
$$L_2
\\=\sum_{i=1}^{n}\sum_{j<k}
\log{f_2(y_{ij},y_{ik};\nu_{ij},\nu_{ik},\gbf,\rho_{jk})},$$
where
$$f_2(y_{ij},y_{ik};\nu_{ij},\nu_{ik},\gbf,\rho_{jk})=\int_{\Phi^{-1}[F_1(y_{ij}-1;\nu_{ij},\gbf)]}^{\Phi^{-1}[F_1(y_{ij};\nu_{ij},\gbf)]}
\int_{\Phi^{-1}[F_1(y_{ik}-1;\nu_{ik}),\gbf]}^{\Phi^{-1}[F_1(y_{ik};\nu_{ik},\gbf)]}  \phi_2(z_j,z_k;\rho_{jk}) dz_j dz_k;
$$
 $\phi_2(\cdot;\rho)$ denotes the standard bivariate normal density  with correlation
 $\rho$.  
Differentiating $L_2$ with respect to $\R$ 
leads to the bivariate composite score function:
\begin{equation}\label{bivariate-CL}
\g_2=\sum_{i=1}^{n}\s_i^{(2)}(\atil,\R)
 =\mathbf{0},
 \end{equation}
where $\s_i^{(2)}(\a,\R)=\frac{\p\sum_{j<k}\log f_2(y_{ij},y_{ik};\nu_{ij},\nu_{ik},\gbf,\rho_{jk})}{\p \R}$.

In what follows, we form weighted versions of the 
CL estimating equations and an iterative approach to determine good weight matrices. 
At the first stage we reform the univariate estimating composite function by inserting weight matrices between the matrix of the covariates $\X_{i}$ and the vector of univariate scores for regression and non-regression  parameters in the univariate composite score function in (\ref{independent score equations}). 
The resulting estimation function is 
\begin{equation}\label{wtee}
  \g_1^\star=\g_1^\star(\a)=\sum_{i=1}^n\X_i^T\,[\W_{i}^{(1)}]^{-1}\,\s_i^{(1)}(\a),
\end{equation}
where  $\W_i^{(1)}$ are invertible $d(1 + q) \times d(1+q)$ matrices.
The estimate of $\a$, denoted by $\hat\a$, is the solution of $\g_1^\star(\a)={\bf 0}$. 

With $\a$ fixed,   as estimated a the first stage of the method, we   also include weight matrices  in the bivariate composite score function in (\ref{bivariate-CL}) at the second stage of the method. 
The resulting estimation function is 
\begin{equation}\label{wtee2}
g_2^\star=g_2^\star(\hat\a,\R)=\sum_{i=1}^{n}[\W_i^{(2)}]^{-1}s_i^{(2)}(\hat\a,\R),
\end{equation}
where  $\W_i^{(2)}$ are invertible $\binom{d}{2} \times \binom{d}{2}$ matrices. We estimate $\R$ by $\hat\R$, a solution of $g_2^\star(\hat\a,\R)=\bf 0$.

The asymptotic covariance matrix for the estimators that solve  the weighted scores estimating equations $\g^\star=(\g_1^\star,\g_2^\star)^\top$,   
viz., 
\begin{equation}\label{inverseGodambe}
\V^\star=(-\bfD_{\g^\star})^{-1}\M_{\g^\star}(-\bfD^T_{\g^\star})^{-1},
\end{equation} 
is used to obtain optimal choices for $\W_i^{(1)}$ and $\W_i^{(2)}$. 
The covariance matrix $\M_\g^\star$  of the estimating  functions $\g^\star$  is 
\begin{eqnarray}\label{J1}
\M_{\g^\star}&=&\mbox{Cov}(\g^\star)=
\begin{pmatrix}\mbox{Cov}(\g_1^\star) & \mbox{Cov}(\g_1^\star,\g_2^\star)\\\mbox{Cov}(\g_2^\star,\g_1^\star) & \mbox{Cov}(\g_2^\star)
\end{pmatrix}
\nonumber\\\nonumber\\
&=&
\begin{pmatrix}\sum_{i=1}^n\X_i^\top[\W_i^{(1)}]^{-1}\O_i^{(1)}[\W_i^{(1)\top}]^{-1}\X_i&&\sum_{i=1}^n \X_i^\top[\W_i^{(1)}]^{-1}\O_i^{(1,2)}[\W_i^{(2)\top}]^{-1}\\\\
\sum_{i=1}^n[\W_i^{(2)}]^{-1}\O_i^{(2,1)}[\W_i^{(1)\top}]^{-1}\X_i&&\sum_{i=1}^n[\W_i^{(2)}]^{-1}\O_i^{(2)}[\W_i^{(2)\top}]^{-1}\end{pmatrix},
\end{eqnarray}
where
$$\begin{pmatrix}\O_i^{(1)}& \O_i^{(1,2)}\\
\O_i^{(2,1)}& \O_i^{(2)}\end{pmatrix}=
\begin{pmatrix}\mbox{Cov}\Bigl(\s_i^{(1)}(\a)\Bigr) & \mbox{Cov}\Bigl(\s_i^{(1)}(\a),\s_i^{(2)}(\a,\R)\Bigr)\\
\mbox{Cov}\Bigl(\s_i^{(2)}(\a,\R),\s_i^{(1)}(\a)\Bigr) & \mbox{Cov}\Bigl(\s_i^{(2)}(\a,\R)\Bigr)
\end{pmatrix}.$$
The Hessian matrix $-\bfD_{\g^\star}$   of the estimating  functions $\g^\star$  is
\begin{eqnarray}\label{D1}
-\bfD_{\g^\star}&=&E\Bigl(\frac{\p \g^\star}{\p\thbf}\Bigr)
=
\begin{pmatrix}
E\Bigl(\frac{\p \g_1^\star}{\p\a}\Bigr)&E\Bigl(\frac{\p \g_1^\star}{\p\R}\Bigr)\\
E\Bigl(\frac{\p \g_2^\star}{\p\a}\Bigr)&E\Bigl(\frac{\p \g_2^\star}{\p\R}\Bigr)
\end{pmatrix}
=
\begin{pmatrix}
-\bfD_{\g_1^\star}&\mathbf{0}\\
-\bfD_{\g_{2,1}^\star}&-\bfD_{\g_2^\star}
\end{pmatrix}\nonumber\\\nonumber\\ 
&=&
\begin{pmatrix}
\sum_{i=1}^n\X_i^\top[\W_i^{(1)}]^{-1}\bfPsi_i^{(1)}&0\\\\
\sum_i^n[\W_i^{(2)}]^{-1}\bfPsi_i^{(2,1)}&\sum_i^n[\W_i^{(2)}]^{-1}\Debf_i^{(2)}\end{pmatrix},
\end{eqnarray}
where $\bfPsi_i^{(1)}=E\Bigl(\frac{\p\s_i^{(1)}(\a)}{\p\a}\Bigr)$, $\bfPsi_i^{(2,1)}=E\Bigl(\frac{\p s_i^{(2)}(\a,\R)}{\p\a}\Bigr)$ and $\Debf_i^{(2)}=E\Bigl(\frac{\p s_i^{(2)}(\a,\R)}{\p\R}\Bigr)$.

The matrix Cauchy-Schwarz inequality  \citep{chaganty97b},
shows that the  optimal choices of $\W_i^{(1)}$ and $\W_i^{(2)}$ satisfy
$\X_i^\top[\W_i^{(1)}]^{-1}=\bfPsi_i^{(1)\top}[\O_i^{(1)}]^{-1}$ and $[\W_i^{(2)}]^{-1}=\Debf_i^{(2)}[\O_i^{(2)}]^{-1}$, respectively, leading to

\begin{equation}\label{J2}
\M_{\g^\star}=
\begin{pmatrix}\sum_{i=1}^n\bfPsi_i^{(1)\top}[\O_i^{(1)}]^{-1}\bfPsi_i^{(1)}&&\sum_{i=1}^n\bfPsi_i^{(1)\top}[\O_i^{(1)}]^{-1}\O_i^{(1,2)}\Debf_i^{(2)}[\O_i^{(2)}]^{-1}\\\\
\sum_{i=1}^n\Debf_i^{(2)}[\O_i^{(2)}]^{-1}\O_i^{(2,1)}[\O_i^{(1)}]^{-1}\bfPsi_i^{(1)}&&\sum_{i=1}^n\Debf_i^{(2)}[\O_i^{(2)}]^{-1}\Debf_i^{(2)}.\end{pmatrix}
\end{equation}
and 
\begin{equation}\label{D2}
-\bfD_{\g^\star}=
\begin{pmatrix}
\sum_{i=1}^n\bfPsi_i^{(1)\top}[\O_i^{(1)}]^{-1}\bfPsi_i^{(1)}&0\\\\
\sum_{i=1}^n\Debf_i^{(2)}[\O_i^{(2)}]^{-1}\bfPsi_i^{(2,1)}&\sum_{i=1}^n\Debf_i^{(2)}[\O_i^{(2)}]^{-1}\Debf_i^{(2)}\end{pmatrix}.
\end{equation}

\bigskip

A few comments on numerical aspects are in order. For an unstructured  dependence  the calculation of $\O_i^{(1)}$ requires summations over the $K^2$ possible cases, while the calculation of $\O_i^{(1,2)}$ and $\O_i^{(2)}$ requires summations up to  
$K^3$ and $K^4$ possible cases, respectively.

\section{\label{asymptotics-sec}Relative efficiency: comparison based on
asymptotic variances}

We consider the MVN copula  with positive exchangeable dependence  and  univariate ordinal logistic regressions as the marginals. For positive exchangeable dependence, if one computes the rectangle MVN probabilities with the 1-dimensional integral method in \cite{Johnson&Kotz72}, then one is using a numerically accurate maximum likelihood (ML)  method that is valid for any dimension \citep{nikoloulopoulos13b,nikoloulopoulos2015a}.

For the covariates, regression and not regression parameters we chose  $p=1, \x_{ij}=x_{1ij}^\top$ where $x_{1ij}$ are taken as uniform random
variables in the interval $[-1,1]$; $\beta_1=0.5$ and $\gbf=(0.33, 0.67)$. 
For the  above  parameters 
we computed the  inverse of the Fisher information matrix $\mathcal{I}$, viz. 
$$\mathcal{I}=\frac{1}{n}\sum_{i=1}^n\sum_{\y_i}
 \frac{\partial f_d(\mathbf{y}_i;\thbf)}{\partial \thbf}
  \frac{\partial f_d(\mathbf{y}_i;\thbf)}{\partial \thbf^{T}}\Bigm/f_d(\mathbf{y}_i;\thbf),$$
and  the matrix
$\V^\star$
with $\M_{\g^\star}$ and $-\bfD_{\g^\star}$ as given in (\ref{J2}) and (\ref{D2}), respectively. The former
 is the asymptotic covariance matrix of the ML estimates, while the latter is the asymptotic covariance matrix of the  WCL estimates of univariate and correlation parameters that are the solutions of the the weighted versions of the CL estimating equations in (\ref{wtee}) and (\ref{wtee2}). We have also computed the asymptotic covariance matrix of the CL estimates, that is the matrix
$\V^\star$
with $\M_{\g^\star}$ and $-\bfD_{\g^\star}$ as given in (\ref{J1}) and (\ref{D1}), respectively, where the weight matrices are simply the identity matrices.

Representative summaries of findings on the
performance and the comparison  of the competing methods
are given in Table \ref{asymVres} 
for three-, six- and nine-dimensional MVN copula models with univariate ordinal logistic regressions. We took
$n=500$ to get a good approximation of the asymptotic efficiency.
The comparisons are made on the scaled diagonal elements,
corresponding  to the asymptotic
variances of the   parameters,
 of the three matrices  with different values of $\rho$.

\begin{table}[!h]
  \centering
   \spacingset{1.}
  \caption{\label{asymVres}Asymptotic variances, scaled by $n$, of the ML, WCL and CL estimates of the MVN copula model parameters. 
Efficiencies with respect to ML are shown in parenthesis.}
  \begin{small}
  \setlength{\tabcolsep}{10pt}
 
   \begin{tabular}{ccccccc}
 \toprule
$d$ & True $\rho$& Method & $\beta_1$ & $\gamma_1$ & $\gamma_2$ &$\rho$\\
\hline
\rowcolor{gray} 3&0.1&ML&4.064 (1.000)&1.572 (1.000)&1.695 (1.000)&0.856 (1.000)\\
\rowcolor{gray} &&WCL&4.064 (1.000)&1.572 (1.000)&1.695 (1.000)&0.856 (1.000)\\
\rowcolor{gray}&&CL&4.097 (0.992)&1.572 (1.000)&1.695 (1.000)&0.856 (1.000)\\
&0.4&ML&3.610 (1.000)&2.107 (1.000)&2.263 (1.000)&0.851 (1.000)\\
 &&WCL&3.623 (0.996)&2.111 (0.998)&2.266 (0.999)&0.856 (0.994)\\
&&CL&4.086 (0.884)&2.112 (0.998)&2.266 (0.999)&0.856 (0.994)\\
\rowcolor{gray}&0.7&ML&2.699 (1.000)&2.711 (1.000)&2.921 (1.000)&0.473 (1.000)\\
\rowcolor{gray} &&WCL&2.744 (0.983)&2.734 (0.992)&2.939 (0.994)&0.481 (0.982)\\
\rowcolor{gray}&&CL&4.104 (0.658)&2.735 (0.991)&2.940 (0.994)&0.482 (0.981)\\
&0.9&ML&1.735 (1.000)&3.241 (1.000)&3.508 (1.000)&0.113 (1.000)\\
 &&WCL&1.866 (0.930)&3.237 (1.001)&3.481 (1.008)&0.116 (0.977)\\
&&CL&4.141 (0.419)&3.24 (1.000)&3.485 (1.007)&0.117 (0.964)\\
\hline
\rowcolor{gray}6&0.1&ML&2.010 (1.000)&0.916 (1.000)&0.983 (1.000)&0.234 (1.000)\\
\rowcolor{gray} &&WCL&2.010 (1.000)&0.916 (1.000)&0.982 (1.000)&0.235 (0.999)\\
\rowcolor{gray}&&CL&2.047 (0.982)&0.916 (1.000)&0.982 (1.000)&0.235 (0.999)\\
&0.4&ML&1.705 (1.000)&1.582 (1.000)&1.69 (1.000)&0.374 (1.000)\\
 &&WCL&1.715 (0.994)&1.591 (0.994)&1.696 (0.996)&0.384 (0.974)\\
&&CL&2.102 (0.811)&1.591 (0.994)&1.696 (0.996)&0.384 (0.974)\\
\rowcolor{gray}&0.7&ML&1.239 (1.000)&2.319 (1.000)&2.494 (1.000)&0.258 (1.000)\\
\rowcolor{gray} &&WCL&1.267 (0.978)&2.37 (0.979)&2.538 (0.983)&0.279 (0.926)\\
\rowcolor{gray}&&CL&2.203 (0.562)&2.371 (0.978)&2.539 (0.982)&0.279 (0.926)\\
&0.9&ML&0.866 (1.000)&2.956 (1.000)&3.205 (1.000)&0.067 (1.000)\\
 &&WCL&0.894 (0.969)&2.999 (0.986)&3.218 (0.996)&0.075 (0.898)\\
&&CL&2.312 (0.375)&3.003 (0.984)&3.221 (0.995)&0.076 (0.889)\\\hline
\rowcolor{gray} 9&0.1&ML&1.327 (1.000)&0.698 (1.000)&0.745 (1.000)&0.128 (1.000)\\
\rowcolor{gray} &&WCL&1.327 (1.000)&0.697 (1.000)&0.745 (1.000)&0.128 (0.999)\\
\rowcolor{gray}&&CL&1.359 (0.976)&0.697 (1.000)&0.745 (1.000)&0.128 (0.999)\\
&0.4&ML&1.096 (1.000)&1.405 (1.000)&1.497 (1.000)&0.272 (1.000)\\
 &&WCL&1.104 (0.993)&1.417 (0.992)&1.506 (0.994)&0.284 (0.957)\\
&&CL&1.382 (0.793)&1.418 (0.991)&1.507 (0.994)&0.284 (0.957)\\
\rowcolor{gray}&0.7&ML&0.806 (1.000)&2.18 (1.000)&2.343 (1.000)&0.204 (1.000)\\
\rowcolor{gray} &&WCL&0.827 (0.975)&2.248 (0.969)&2.404 (0.974)&0.231 (0.884)\\
\rowcolor{gray}&&CL&1.448 (0.557)&2.249 (0.969)&2.405 (0.974)&0.231 (0.884)\\
&0.9&ML&0.621 (1.000)&2.843 (1.000)&3.085 (1.000)&0.055 (1.000)\\
 &&WCL&0.621 (1.001)&2.919 (0.974)&3.130 (0.986)&0.065 (0.844)\\
&&CL&1.530 (0.406)&2.923 (0.973)&3.133 (0.985)&0.065 (0.838)\\
\bottomrule
 \end{tabular}%
 \end{small}
\end{table}

Conclusions from the values in the table
 are the following.

\begin{itemize}

\itemsep=10pt
\item The  WCL method for the univariate parameters 
 is nearly as efficient as ML. 
 
 \item The  CL method for the univariate parameters 
 is inefficient as the asymptotic variances are overestimated.  
 
 \item The  WCL method for the correlation parameters shares similar efficiency with the  CL method. That is, their efficiency decreases for strong correlation as the dimension increases. This is not though a worry as  for real discrete response data one does not except correlations greater than 0.8.

\end{itemize}

\section{\label{sim-sec}Simulations}

We study the small-sample efficiency of the weighted composite score functions in both a low- and high-dimensional case.
Section \ref{longitudinal-sim-sec}   focuses on  simulated longitudinal ordinal data with small size clusters. Section \ref{ts-sim} contains simulations for ordinal time-series of dimension up to 1000.

\subsection{\label{longitudinal-sim-sec}Longitudinal ordinal}

We randomly generate $B=10^4$ samples of size  {$n = 100, 300, 500$} from the   multivariate ordinal probit  model  with  an unstructured dependence. 
We use   the same dimension ($d=4$) and latent correlation matrix as in \cite{pitetal2006}, viz.,  

$$\R=
\begin{pmatrix}
1& 0.6348 & 0.5821& 0.6916\\
0.6348 &1 & 0.3662 & 0.8059\\
0.5821&0.3662& 1&0.0435\\
0.6916&0.8059&0.0435&1
\end{pmatrix}
$$
and $K=5$ ordinal categories (equally weighted). 
For the covariates and ordinal probit regression parameters, we chose $p=4,\x_{ij}=(x_{1ij},x_{2ij},x_{3ij},x_{4ij})^\top$ with $x_{1ij}$ the time,  $x_{2ij}\in\{0,1\}$ a group variable, ,  $x_{3ij}=x_{1ij}\times x_{2ij}$, and $x_{4ij}$
a uniform random
variable in the interval $[-1,1]$; $\beta_1=-\beta_2=-\beta_3=-0.5, \b_4=1$.

Table \ref{longitudinal-sim} contains the parameter values, the bias,
standard deviations (SD) and root mean square errors (RMSE) of the
 WCL and CL estimates, scaled by $n$,  along
with the average of their theoretical SDs (denoted with $\sqrt{\bar V}$). Note in passing, for the weighted scores estimates, $V$ is the diagonal of $\V_1^\star$. 
It is clear from the table that both the WCL and CL methods provide unbiased estimates and variances computed from the simulations are similar to the asymptotic variances for both the WCL and CL methods.

\begin{sidewaystable}[htbp]
 \centering
  \spacingset{1.}
\caption{\label{longitudinal-sim}Small sample of sizes $n = 100, 300, 500$ simulations ($10^4$ replications) and resulted biases, standard deviations (SD), and root mean square errors (RMSE), scaled by $n$, for the WCL and CL eatimates of the correlation parameters for the quadrivariate  MVN copula with an unstructured dependence and ordinal probit regressions.}
 \setlength{\tabcolsep}{6pt}

   \begin{tabular}{llccccccccccccccc}
   \toprule
         & Method & $n$ & $\beta_1$ & $\beta_2$ & $\beta_3$ & $\beta_4$ & $\gamma_1$ & $\gamma_2$ & $\gamma_3$ & $\gamma_4$ & $\rho_{12}$ & $\rho_{13}$ & $\rho_{14}$ & $\rho_{23}$ & $\rho_{24}$ & $\rho_{34}$ \\
   \midrule
   $n$Bias  & CL    & 100   & 0.65  & -0.90 & -0.65 & 1.34  & -0.83 & -0.09 & 0.56  & 1.36  & -0.76 & -1.19 & -0.43 & -1.37 & -0.81 & -1.05 \\
\rowcolor{gray}         & WCL    &       & 0.61  & -0.95 & -0.60 & 1.44  & -0.81 & -0.05 & 0.62  & 1.41  & -1.10 & -1.24 & -0.66 & -1.08 & -0.55 & 0.13 \\
         & CL    & 300   & 0.66  & -1.12 & -0.60 & 1.49  & -0.98 & -0.19 & 0.40  & 1.14  & -0.88 & -1.24 & -0.32 & -1.68 & -0.70 & -1.48 \\
\rowcolor{gray}         & WCL    &       & 0.58  & -1.10 & -0.53 & 1.45  & -0.97 & -0.14 & 0.48  & 1.20  & -1.08 & -1.16 & -0.50 & -1.48 & -0.45 & -0.99 \\
         & CL    & 500   & 0.79  & -0.10 & -0.81 & 1.00  & -1.64 & -0.60 & 0.03  & 0.76  & -1.10 & -1.31 & -0.59 & -1.49 & -0.84 & -1.83 \\
\rowcolor{gray}         & WCL    &       & 0.72  & -0.10 & -0.70 & 1.25  & -1.66 & -0.64 & 0.01  & 0.76  & -1.16 & -1.09 & -0.65 & -1.17 & -0.46 & -1.60 \\ \midrule
   $n$SD    & CL    & 100   & 5.16  & 20.34 & 6.26  & 11.36 & 16.05 & 15.13 & 15.07 & 15.72 & 7.80  & 8.56  & 7.32  & 11.20 & 5.33  & 13.65 \\
 \rowcolor{gray}        & WCL    &       & 4.76  & 20.10 & 5.71  & 8.88  & 15.89 & 14.98 & 14.91 & 15.53 & 8.82  & 9.59  & 9.47  & 12.05 & 7.77  & 14.82 \\
         & CL    & 300   & 8.83  & 35.47 & 10.80 & 19.43 & 27.48 & 26.03 & 25.85 & 27.07 & 13.45 & 14.71 & 12.52 & 19.34 & 8.81  & 23.77 \\
\rowcolor{gray}         & WCL    &       & 7.99  & 34.98 & 9.76  & 14.99 & 27.23 & 25.76 & 25.57 & 26.75 & 15.17 & 15.79 & 14.33 & 20.13 & 11.72 & 25.54 \\
         & CL    & 500   & 11.36 & 45.52 & 13.90 & 24.98 & 35.41 & 33.80 & 33.85 & 35.09 & 17.04 & 19.00 & 16.06 & 24.88 & 11.37 & 30.50 \\
\rowcolor{gray}         & WCL    &       & 10.33 & 44.89 & 12.60 & 19.13 & 35.01 & 33.41 & 33.37 & 34.51 & 18.34 & 19.62 & 16.89 & 25.44 & 12.51 & 32.77 \\ \midrule
   $n\sqrt{\bar V}$ & CL    & 100   & 5.15  & 20.09 & 6.25  & 11.27 & 15.68 & 14.86 & 14.78 & 15.46 & 7.64  & 8.60  & 7.29  & 11.05 & 5.20  & 13.40 \\
\rowcolor{gray}         & WCL    &       & 4.63  & 19.73 & 5.57  & 8.63  & 15.46 & 14.63 & 14.53 & 15.18 & 9.78  & 10.22 & 8.97  & 13.58 & 6.71  & 17.07 \\
         & CL    & 300   & 8.78  & 34.76 & 10.69 & 19.33 & 27.07 & 25.67 & 25.52 & 26.67 & 13.16 & 14.76 & 12.57 & 19.14 & 8.80  & 23.40 \\
\rowcolor{gray}         & WCL    &       & 7.93  & 34.22 & 9.55  & 14.77 & 26.75 & 25.34 & 25.16 & 26.27 & 13.87 & 16.23 & 14.45 & 19.96 & 10.39 & 23.17 \\
         & CL    & 500   & 11.30 & 44.86 & 13.76 & 24.91 & 34.93 & 33.12 & 32.93 & 34.40 & 16.97 & 19.01 & 16.22 & 24.68 & 11.33 & 30.24 \\
 \rowcolor{gray}        & WCL    &       & 10.21 & 44.19 & 12.31 & 19.02 & 34.53 & 32.71 & 32.49 & 33.91 & 17.81 & 20.79 & 18.40 & 25.62 & 13.16 & 29.85 \\ \midrule
   $n$RMSE  & CL    & 100   & 5.20  & 20.36 & 6.30  & 11.44 & 16.08 & 15.13 & 15.08 & 15.78 & 7.83  & 8.65  & 7.33 & 11.28 & 5.39  & 13.69 \\
 \rowcolor{gray}        & WCL    &       & 4.80  & 20.12 & 5.75  & 9.00  & 15.91 & 14.98 & 14.92 & 15.59 & 8.89  & 9.67  & 9.49  & 12.10 & 7.79  & 14.82 \\
         & CL    & 300   & 8.86  & 35.48 & 10.82 & 19.49 & 27.50 & 26.03 & 25.85 & 27.09 & 13.48 & 14.77 & 12.53 & 19.42 & 8.84  & 23.82 \\
 \rowcolor{gray}        & WCL    &       & 8.01  & 34.99 & 9.78  & 15.06 & 27.25 & 25.76 & 25.58 & 26.78 & 15.20 & 15.83 & 14.34 & 20.18 & 11.72 & 25.56 \\
         & CL    & 500   & 11.39 & 45.52 & 13.92 & 25.00 & 35.45 & 33.80 & 33.85 & 35.10 & 17.08 & 19.05 & 16.07 & 24.92 & 11.40 & 30.55 \\
 \rowcolor{gray}        & WCL    &       & 10.36 & 44.89 & 12.61 & 19.17 & 35.05 & 33.42 & 33.37 & 34.52 & 18.38 & 19.65 & 16.90 & 25.47 & 12.52 & 32.81 \\
   \bottomrule
   \end{tabular}%
 \label{tab:addlabel}%
\end{sidewaystable}

\subsection{\label{ts-sim}Ordinal time series}

We randomly generate $B=10^4$ samples of dimension  $d =100,200, 500,1000$ from the  multivariate ordinal probit model   with  latent correlation matrix  corresponding to that of an autoregressive process of order one with first-order autocorrelation equals to 0.8 and $K=4$ categories. The above problem differs from the simulations in the preceding section in that rather than being repeated or clustered measurements of a variable, the data are multivariate, with a single measurement on each of $d$ different variables.

For the covariates and regression parameters, we use a combination of a time-stationary and a time-varying design, i.e., include 
covariates that are typically constant over time,  and correlated over time. 
More specifically, we chose $p=4,\x_{ij}=(x_{1ij},x_{2ij},x_{3ij})^\top$ 
with $x_{1i}\in\{0,1\}$ a binary variable with probability of success 0.4, $x_{2ij}$ a time-varying variable from a $d$-variate MVN copula with  latent correlation matrix  corresponding to that of an autoregressive process of order one with first-order autocorrelation equals to 0.5,  $x_{3ij}=x_{1ij}\times x_{2ij}$; 
$\beta_1=-\beta_2=\beta_3=-0.5$. 

For a structured  dependence and high dimension $d$,   the computation of the bivariate weight matrices involved is prohibitive as the calculation of $\O_i^{(1,2)}$ and $\O_i^{(2)}$
requires summations  over the $K^d$ possible cases. Nevertheless this is not the case for the calculation of $\O_i^{(1)}$ in the univariate weight matrices which only requires summations over the $K^2$ possibles cases. Hence, we study the small-sample efficiency of the WCL estimating equations for the univariate regression or non-regression parameters, i.e., the first stage of the proposed WCL method.

Table \ref{tseries-sim} contains the parameter values, the bias,
standard deviations (SD) and root mean square errors (RMSE) of the
 WCL and CL estimates, scaled by $d$,  along
with the average of their theoretical SDs (denoted with $\sqrt{\bar V}$). 
It is clear from the table that both the WCL and CL methods provide unbiased estimates. 
The variances computed from the simulations are similar to the asymptotic variances for both the WCL and CL methods. 
The WCL estimates of the regression parameters are remarkably more efficient than the CL estimates as the variances in the CL method are overestimated.

\begin{sidewaystable}[htbp]
 \centering
  \spacingset{1.}
 \caption{\label{tseries-sim}
 Small sample of time-series lengths $d =100,200, 500,1000$  simulations ($10^4$ replications) and resulted biases, standard deviations (SD), and root mean square errors (RMSE), scaled by $d$, 
 for the WCL and CL estimates of the univariate parameters for the  multivariate ordinal probit model   with  latent correlation matrix  corresponding to that of an autoregressive process of order one with first-order autocorrelation equals to 0.8 and $K=4$ categories. 
}

\setlength{\tabcolsep}{21pt}

   \begin{tabular}{rcccccccc}
   \toprule
         & $d$ &       & $\b_1$ & $\b_2$ & $\b_3$ & $\gamma_1$ & $\gamma_2$ & $\gamma_3$ \\
   \midrule
   $d$Bias  & 100   & CL    & 4.97  & -4.37 & 4.79  & -6.24 & 0.02  & 6.26 \\
 \rowcolor{gray}         &       & WCL   & 5.04  & -4.72 & 5.17  & -6.34 & 0.20  & 6.58 \\
         & 200   & CL    & 5.10  & -5.12 & 4.59  & -5.92 & 0.07  & 5.76 \\
\rowcolor{gray}          &       & WCL   & 5.33  & -4.94 & 3.94  & -6.16 & 0.26  & 6.21 \\
         & 500   & CL    & 5.41  & -6.01 & 5.28  & -4.96 & 0.80  & 6.94 \\
\rowcolor{gray}          &       & WCL   & 4.69  & -5.90 & 5.73  & -5.13 & 1.09  & 7.52 \\
         & 1000  & CL    & 5.65  & -6.97 & 3.94  & -5.41 & 0.97  & 7.11 \\
\rowcolor{gray}          &       & WCL   & 4.48  & -6.62 & 7.07  & -5.87 & 1.02  & 7.31 \\
   $d$SD    & 100   & CL    & 49.46 & 68.20 & 86.34 & 49.89 & 47.99 & 49.57 \\
 \rowcolor{gray}         &       & WCL   & 32.33 & 41.54 & 55.39 & 41.02 & 38.72 & 40.77 \\
         & 200   & CL    & 68.01 & 93.52 & 117.17 & 68.08 & 66.27 & 68.47 \\
 \rowcolor{gray}         &       & WCL   & 42.68 & 53.91 & 72.96 & 54.90 & 52.53 & 55.24 \\
         & 500   & CL    & 103.72 & 145.21 & 180.47 & 106.17 & 103.41 & 106.56 \\
 \rowcolor{gray}         &       & WCL   & 63.16 & 81.27 & 108.30 & 85.81 & 82.04 & 86.03 \\
         & 1000  & CL    & 144.65 & 202.00 & 250.58 & 148.67 & 144.94 & 148.27 \\
 \rowcolor{gray}         &       & WCL   & 87.57 & 110.95 & 150.69 & 118.98 & 113.92 & 118.54 \\
   $d\sqrt{V}$ & 100   & CL    & 46.49 & 61.65 & 81.00 & 43.57 & 42.28 & 43.56 \\
  \rowcolor{gray}        &       & WCL   & 32.45 & 40.61 & 56.22 & 36.12 & 34.52 & 36.11 \\
         & 200   & CL    & 64.98 & 88.27 & 112.58 & 63.81 & 62.00 & 63.69 \\
 \rowcolor{gray}         &       & WCL   & 42.31 & 53.43 & 72.61 & 52.15 & 49.80 & 52.04 \\
         & 500   & CL    & 102.43 & 141.11 & 176.85 & 103.22 & 100.27 & 103.04 \\
 \rowcolor{gray}         &       & WCL   & 63.81 & 80.95 & 108.88 & 83.76 & 79.86 & 83.60 \\
         & 1000  & CL    & 144.93 & 200.68 & 249.90 & 147.33 & 142.78 & 147.01 \\
 \rowcolor{gray}         &       & WCL   & 89.10 & 113.14 & 151.54 & 119.42 & 113.41 & 119.11 \\
   $d$RMSE  & 100   & CL    & 49.71 & 68.34 & 86.47 & 50.28 & 47.99 & 49.96 \\
 \rowcolor{gray}         &       & WCL   & 32.72 & 41.81 & 55.63 & 41.50 & 38.72 & 41.29 \\
         & 200   & CL    & 68.21 & 93.66 & 117.26 & 68.34 & 66.27 & 68.71 \\
 \rowcolor{gray}         &       & WCL   & 43.01 & 54.14 & 73.06 & 55.25 & 52.53 & 55.58 \\
         & 500   & CL    & 103.86 & 145.33 & 180.55 & 106.29 & 103.41 & 106.78 \\
 \rowcolor{gray}         &       & WCL   & 63.34 & 81.48 & 108.45 & 85.96 & 82.04 & 86.36 \\
         & 1000  & CL    & 144.76 & 202.12 & 250.61 & 148.77 & 144.94 & 148.44 \\
  \rowcolor{gray}        &       & WCL   & 87.69 & 111.14 & 150.86 & 119.12 & 113.92 & 118.77 \\
   \bottomrule
   \end{tabular}%
\end{sidewaystable}

\section{\label{sec-app}Applications}

In this section we illustrate the proposed estimation method through two examples
with longitudinal (low-dimensional) and time (high-dimensional) series ordinal response data. 
We include  comparisons with other possible  fitting methods, such as ML (low dimension) and CL (high dimension). 
\subsection{Arthritis data}

We illustrate the  weighted composite scores equations  by analysing the rheumatoid arthritis data-set \citep{Bombardier-etal-1986}.  
The data were taken from a randomized clinical trial  designed to evaluate  the effectiveness of the treatment Auranofin versus a placebo therapy for the treatment of rheumatoid arthritis. The repeated ordinal response is the self-assessment of arthritis,  classified on a five-level ordinal scale $(1 = \mbox{poor},\ldots, 5 = \mbox{very good})$. Patients ($n=303$) were randomized into one of the two treatment groups after baseline self-assessment followed during five months  of treatment with  measurements 
taken at the  first month and  every two months during treatment resulting in a maximum of 3 measurements per subject (unequal cluster sizes). The  covariates are time, baseline-assessment, age in years at baseline, sex and treatment. We treat time and baseline-assessment as categorical variables  
and look at differences between adjacent outcome categories (see, e.g., \citealp{Tutz&Gertheiss2016}). To this end we followed the coding scheme for ordinal independent variables in \cite{Walter-etal-1987}. 
Further, both logit and probit links are used for the ordinal regressions.

Table \ref{arthritis} gives the estimates  and standard errors  of the
model parameters obtained using ML and the  WCL estimating equations. 
Ordinal logistic regression is slightly better than ordinal probit regression based on the composite and full likelihoods. 
Our analysis shows
 that the  WCL estimates of all the parameters and their corresponding standard errors 
 are  nearly the same as the ML estimates.

\begin{table}[!h]
 \centering
  \spacingset{1.}
 \caption{\label{arthritis} WCL and ML estimates (Est.) along with their standard errors (SE) for the arthritis data.}
 \setlength{\tabcolsep}{6pt}

   \begin{tabular}{ccccccccccccc}
   \toprule
         &       & \multicolumn{5}{c}{Logit link}        &       & \multicolumn{5}{c}{Probit link} \\ 
\cmidrule{1-1}         \cmidrule{3-7}  \cmidrule{9-13}
 Covariates \&         &       & \multicolumn{2}{c}{WCL} &       & \multicolumn{2}{c}{ML} &       & \multicolumn{2}{c}{WCL} &       & \multicolumn{2}{c}{ML} \\
cutpoints         &       & Est.  & SE    &       & Est.  & SE    &       & Est.  & SE    &       & Est.  & SE \\\cmidrule{1-1}  \cmidrule{3-4} \cmidrule{6-7} \cmidrule{9-10} \cmidrule{12-13}
   $I(\mbox{time}=2)$ &       & -0.007 & 0.124 &       & -0.006 & 0.125 &       & -0.005 & 0.071 &      & -0.007 & 0.072 \\
   $I(\mbox{time}=3)$ &       & -0.377 & 0.116 &       & -0.377 & 0.115 &       & -0.218 & 0.066 &       & -0.220 & 0.066 \\
   trt   &       & -0.500 & 0.168 &       & -0.487 & 0.165 &       & -0.337 & 0.097 &       & -0.336 & 0.097 \\
   $I(\mbox{baseline}=2)$ &       & -0.659 & 0.345 &       & -0.607 & 0.357 &       & -0.336 & 0.200 &       & -0.341 & 0.200 \\
   $I(\mbox{baseline}=3)$ &       & -1.208 & 0.329 &       & -1.161 & 0.337 &       & -0.580 & 0.190 &       & -0.576 & 0.190 \\
   $I(\mbox{baseline}=4)$ &       & -2.569 & 0.370 &       & -2.487 & 0.382 &       & -1.319 & 0.211 &       & -1.315 & 0.211 \\
   $I(\mbox{baseline}=5)$ &       & -4.040 & 0.555 &       & -3.975 & 0.549 &       & -2.264 & 0.324 &       & -2.262 & 0.320 \\
   age   &       & 0.013 & 0.008 &       & 0.014 & 0.007 &       & 0.008 & 0.004 &       & 0.008 & 0.004 \\
   sex   &       & -0.167 & 0.187 &       & -0.179 & 0.179 &       & -0.062 & 0.109 &       & -0.062 & 0.108 \\
   $\alpha_1$ &       & -1.768 & 0.673 &       & -1.831 & 0.625 &       & -1.029 & 0.385 &       & -1.016 & 0.383 \\
   $\alpha_2$ &       & 0.351 & 0.656 &       & 0.230 & 0.608 &       & 0.071 & 0.381 &       & 0.059 & 0.382 \\
   $\alpha_3$ &       & 2.324 & 0.662 &       & 2.222 & 0.613 &       & 1.249 & 0.383 &       & 1.250 & 0.385 \\
   $\alpha_4$ &       & 4.641 & 0.682 &       & 4.526 & 0.625 &       & 2.544 & 0.390 &       & 2.545 & 0.390 \\
   $\rho_{12}$ &       & 0.393 & 0.057 &       & 0.376 & 0.061 &       & 0.393 & 0.057 &       & 0.373 & 0.061 \\
   $\rho_{13}$ &       & 0.505 & 0.051 &       & 0.503 & 0.052 &       & 0.509 & 0.051 &       & 0.505 & 0.052 \\
   $\rho_{23}$ &       & 0.530 & 0.050 &       & 0.536 & 0.046 &       & 0.523 & 0.050 &       & 0.528 & 0.046 \\\hline
   Log-likelihood &       &  \multicolumn{2}{c}{-2114.855}       &       & \multicolumn{2}{c}{-1041.477}        &       &      \multicolumn{2}{c}{-2117.755} & & \multicolumn{2}{c}{-1043.083}   \\
   \bottomrule
   \end{tabular}%
 \label{tab:addlabel}%
\end{table}

\subsection{Infant sleep status data}
The sleep data (e.g., \citealt{Fokianos&Kedem-2003-StaSci}) consist of sleep state measurements of a newborn infant together with his heart rate and temperature  sampled every 30 seconds. The sleep states are classified as:
(1) quiet sleep,
(2) indeterminate sleep, (3) active sleep,
(4) awake.
The total number of observations is equal to 1024 and the objective is to predict  the sleep state based on covariate information.

The  response, sleep state, is an ordinal time series in the sense that  the response increases from awake to active sleep, i.e., “(4)” $<$ “(1)” $<$ “(2)” $<$ “(3)”. We use the standardized heart rate and temperature as covariates to avoid large estimates for the univariate cutpoints. \cite{Fokianos&Kedem-2003-StaSci} have previously adopted regression models for this  ordinal time series and confirmed an autoregressive model of order 1 to adequately capture the serial dependence among the ordinal observations.

\begin{table}[!h]
 \centering
  \spacingset{1.}
 \caption{\label{sleep}WCL and CL estimates (Est.) along with their standard errors (SE) for the sleep data.}
 \setlength{\tabcolsep}{7pt}

   \begin{tabular}{ccccccccccc}
   \toprule
               \multicolumn{11}{l}{Logit link} \\
\midrule
   Covariates \&      &       & \multicolumn{4}{c}{CL} & \multicolumn{1}{c}{} & \multicolumn{4}{c}{WCL} \\
      
cutpoints         &       & \multicolumn{1}{c}{Est. } & \multicolumn{1}{c}{SE} & \multicolumn{1}{c}{$Z$} & \multicolumn{1}{c}{$p$-value} & \multicolumn{1}{c}{} & \multicolumn{1}{c}{Est. } & \multicolumn{1}{c}{SE} & \multicolumn{1}{c}{$Z$} & \multicolumn{1}{c}{$p$-value} \\
\cmidrule{1-1}            \cmidrule{3-6} \cmidrule{8-11}
   heart rate &       & 0.074 & 0.274 & 0.271 & 0.786 &       & 0.097 & 0.044 & 2.232 & 0.026 \\
   temperature &       & 0.284 & 0.339 & 0.837 & 0.402 &       & 0.254 & 0.190 & 1.336 & 0.182 \\
   $\alpha_1$ &       & -0.934 & 0.385 & -2.425 & 0.015 &       & -0.778 & 0.372 & -2.091 & 0.037 \\
   $\alpha_2$ &       & 0.771 & 0.376 & 2.053 & 0.040 &       & 0.865 & 0.376 & 2.300 & 0.021 \\
   $\alpha_3$ &       & 1.233 & 0.406 & 3.040 & 0.002 &       & 1.322 & 0.408 & 3.240 & 0.001 \\\hline
     $L_2$    &       & \multicolumn{9}{c}{-1321423} \\
\midrule
       \multicolumn{11}{l}{Probit link} \\
   \midrule
  Covariates \&       &       & \multicolumn{4}{c}{CL} & \multicolumn{1}{c}{} & \multicolumn{4}{c}{WCL} \\
 cutpoints        &       & \multicolumn{1}{c}{Est. } & \multicolumn{1}{c}{SE} & \multicolumn{1}{c}{$Z$} & \multicolumn{1}{c}{$p$-value} & \multicolumn{1}{c}{} & \multicolumn{1}{c}{Est. } & \multicolumn{1}{c}{SE} & \multicolumn{1}{c}{$Z$} & \multicolumn{1}{c}{$p$-value} \\
   \cmidrule{1-1}        \cmidrule{3-6} \cmidrule{8-11}
   heart rate &       & 0.040 & 0.164 & 0.245 & 0.807 &       & 0.057 & 0.026 & 2.221 & 0.026 \\
   temperature &       & 0.167 & 0.202 & 0.827 & 0.408 &       & 0.169 & 0.114 & 1.479 & 0.139 \\
   $\alpha_1$ &       & -0.582 & 0.231 & -2.520 & 0.012 &       & -0.515 & 0.227 & -2.268 & 0.023 \\
   $\alpha_2$ &       & 0.469 & 0.228 & 2.058 & 0.040 &       & 0.510 & 0.227 & 2.250 & 0.024 \\
   $\alpha_3$ &       & 0.744 & 0.236 & 3.154 & 0.002 &       & 0.781 & 0.235 & 3.317 & 0.001 \\\hline
$L_2$  &       & \multicolumn{9}{c}{-1321157}\\
   \bottomrule
   \end{tabular}%
 \label{tab:addlabel}%
\end{table}

Table \ref{sleep} summarizes the WCL and CL estimates of the  regression and non-regression parameters. The first order autoregressive parameter is estimated as $0.96$ for both logistic and probit ordinal regression. The latter is slightly better than ordinal logit regression based on the composite likelihood values. It is obvious from the table that ignoring the actual serial dependence   in the data  on the CL estimation of the regression parameters leads to invalid conclusions resulting to  no effect of the time-dependent covariates  at sleep state. The WCL analyses reveal that the   the heart rate  effect is a statistically significant predictor 
of sleep state.

\section{Discussion}

We have studied a weighted composite likelihood estimating equations approach, namely the WCL,  based on weighting the univariate and bivariate scores of the univariate and bivariate margins of a MVN copula model with discrete margins  to estimate the model parameters. The WCL method leads to efficient estimating equations for both the univariate (regression and non-regression) and correlation parameters.

The proposed method can be used for high dimensional discrete data such as discrete time or spatial series. As such data require a structured correlation, such as an autoregressive moving average or the Mat\'ern isotropic structure,  the second stage of the method, i.e., the estimation of the correlation parameters becomes cumbersome. Nevertheless,  the typical CL estimates of the correlation parameters  can be rather  used  as they share the same efficiency with the WCL correlation estimates. For the first stage of the method  as the weight matrices of the  WCL estimating equations in (\ref{wtee}) depend on covariances of the univariate scores only the 
 bivariate marginal probabilities are needed for estimation. Hence, estimation of MVN copula regression models is feasible for high-dimensional discrete response data with covariates. 

\cite{nikoloulopoulos&joe&chaganty10}  and \cite{Nikoloulopoulos2015d,Nikoloulopoulos-2016-wtsc-ord}
have considered estimating equations of this form when dependence is considered a nuisance; this leads to an extension of generalized estimating equations of \cite{liang&zeger86} in that  a wider class of univariate regression models can be called. 
A sandwich-type estimator is used to obtain estimates of the covariance matrix of model parameters that are robust to misspecification.
Nevertheless, such methods of inference about regression and non-regression parameters are not available in the setting where is a single measurement on each of $d$ different variables.


\end{document}